\documentclass[aps,prl,twocolumn,amsmath,amssymb,superscriptaddress]{revtex4-1}
\usepackage{amsmath}
\usepackage{amssymb}
\usepackage{amsthm}
\usepackage[pdftex]{color}
\usepackage{graphicx}% Include figure files
\usepackage{dcolumn} % Align table columns on decimal point
\usepackage{bm} % bold math
\usepackage{epic}
\usepackage{bbm}
\usepackage{longtable}
\usepackage{slashed}
\usepackage{ulem}   % to strike things out
\begin{document}

\title{Reply to ``Comment on `Floquet Fractional Chern Insulators'"}

\author{Adolfo G. Grushin}
\affiliation{Max-Planck-Institut f\"{u}r Physik komplexer Systeme, 01187 Dresden, Germany}
\author{\'Alvaro G\'omez-Le\'on}
\affiliation{Instituto de Ciencia de Materiales de Madrid, CSIC, Cantoblanco,
E-28049 Madrid, Spain}
\author{Titus Neupert}
\affiliation{Princeton Center for Theoretical Science, Princeton University, Princeton,
New Jersey 08544, USA}

\date{\today}

\begin{abstract}
We respond to the comments expressed by L. D'Alessio in arXiv:1412.3481 on our work 
``Floquet Fractional Chern Insulators" [Phys. Rev. Lett. 112, 156801 (2014)].
We confirm the central result that 
the ground state of the effective Hamiltonian is an
interacting fractional Chern insulator.
\end{abstract}

\maketitle

In this Reply we address the two points of criticism regarding our Letter~\cite{GGN14} that have been expressed first in Ref.~\cite{BDP14} and restated in Ref.~\cite{D14}. These are:
i) The calculation of the Floquet Hamiltonian of the system driven with frequency $\omega$ misses order $\omega^{-1}$ terms and 
ii) the assumption that the Floquet bands are filled as in time-independent systems is questionable.
\\

Let us start with the response to statement i). We are interested in the effective Hamiltonian governing the time evolution 
of a periodically driven honeycomb lattice system of the form $\hat{H}(t)=\hat{H}_{\text{hop}}(t)+\hat{H}_{\text{int}}$. Here $\hat{H}_{\text{hop}}(t)$ includes the
single particle Hamiltonian and the periodic driving and $\hat{H}_{\text{int}}$ is the interacting Hamiltonian, both defined in the first equation of~\cite{GGN14}.
The stroboscopic evolution operator that governs the dynamics of a given initial state is, without loss of generality~\cite{GD14}
\begin{equation}
\label{eq:evol}
\hat{U}(T+\tau, \tau) = e^{i \hat{K}(\tau)} e^{-i \hat{H}_\text{eff} T} e^{-i \hat{K}(\tau)} ,
\end{equation}
where $T$ is the period of the driving.
It is written in terms of a time periodic kick operator $\hat{K}(t)=\hat{K}(t+T)$, that encodes all dependence on initial conditions through $\tau$ that averages to zero over one period,
and defines the $\tau$-independent effective Hamiltonian $\hat{H}_{\text{eff}}$.
In the limit where the driving frequency $\omega=2\pi/T$ is large compared to the other characteristic energy scales of the problem,
the effective Hamiltonian takes the form (using notation of Refs.~\cite{JMD14,GD14})
\begin{subequations}
\label{eq:tauindep}
\begin{align}
\hat{H}_{\text{eff}} &= \hat{H}_{0\omega} + \hat{H}_{1\omega}  + \mathcal{O}\left(\frac{1}{\omega^2}\right), \\
\text{with} \quad \hat{H}_{0\omega} &= \hat{H}_0 \label{eq:order0noT} ,\\
\hat{H}_{1\omega} &= \frac{1}{\omega} \sum_{n = 1}^{\infty} \frac{1}{n} [\hat{H}_n, \hat{H}_{-n}] ,
\label{eq:order1noT}
\end{align}
\end{subequations}
up to first order in $\omega^{-1}$, where the components $\hat{H}_{n}$ define the Fourier transform $\hat{H}(t) = \sum_{n=-\infty}^{+\infty} \hat{H}_n e^{i n \omega t}$.
Two essential properties of $H_{\text{eff}}$ to order $\omega^{-1}$ are relevant for our discussion: a) It only includes $[\hat{H}_n, \hat{H}_{-n}]$ commutators.
 b) Since the interaction term $\hat{H}_{\text{int}}$ is time-independent in our problem, to order $\omega^{-1}$ the only term in which the interaction enters is $\hat{H}_{0\omega}=\hat{H}_{0}$. Therefore, \emph{there is no interaction correction} up to order $\omega^{-1}$ to the effective Hamiltonian $\hat{H}_\text{eff}$ generically~\cite{RGF03}. \\
 
Instead of separating the stroboscopic time evolution operator as done in Eq.~\eqref{eq:evol}, we could have chosen to directly define an evolution operator $\hat{H}_{\text{F}}(\tau)$, the so-called Floquet Hamiltonian, by writing
\begin{equation}
\hat{U}(T+\tau, \tau) = 
e^{-i \hat{H}_{\text{F}}(\tau)T}  ,
\end{equation}
at the expense of an explicit dependence of $\hat{H}_{\text{F}}(\tau)$ on the initial time $\tau$.
The Floquet Hamiltonian $\hat{H}_{\text{F}}(\tau)$ reads to order $\omega^{-1}$
\begin{equation}
\hat{H}_{\text{F}}(\tau) = \hat{H}_{0\omega} + \hat{H}_{1\omega}^\tau + \mathcal{O}\left(\frac{1}{\omega^2}\right),
\end{equation}
with
\begin{subequations}
\label{eq:om}
\begin{eqnarray}
\label{eq:om0}
\hat{H}_{0\omega} &=& \hat{H}_0,\\
\label{eq:om1}
\hat{H}_{1\omega}^\tau &=& \frac{1}{\omega} \sum_{n = 1}^{\infty} \frac{1}{n} \Big([\hat{H}_n, \hat{H}_{-n}] 
\nonumber\\
&&
+ e^{- i n \omega \tau} [\hat{H}_{- n}, \hat{H}_0]
- e^{ i n \omega \tau} [\hat{H}_{+n}, \hat{H}_0]
 \Big),
\end{eqnarray}
\end{subequations}
\begin{figure}
\includegraphics[scale=0.42]{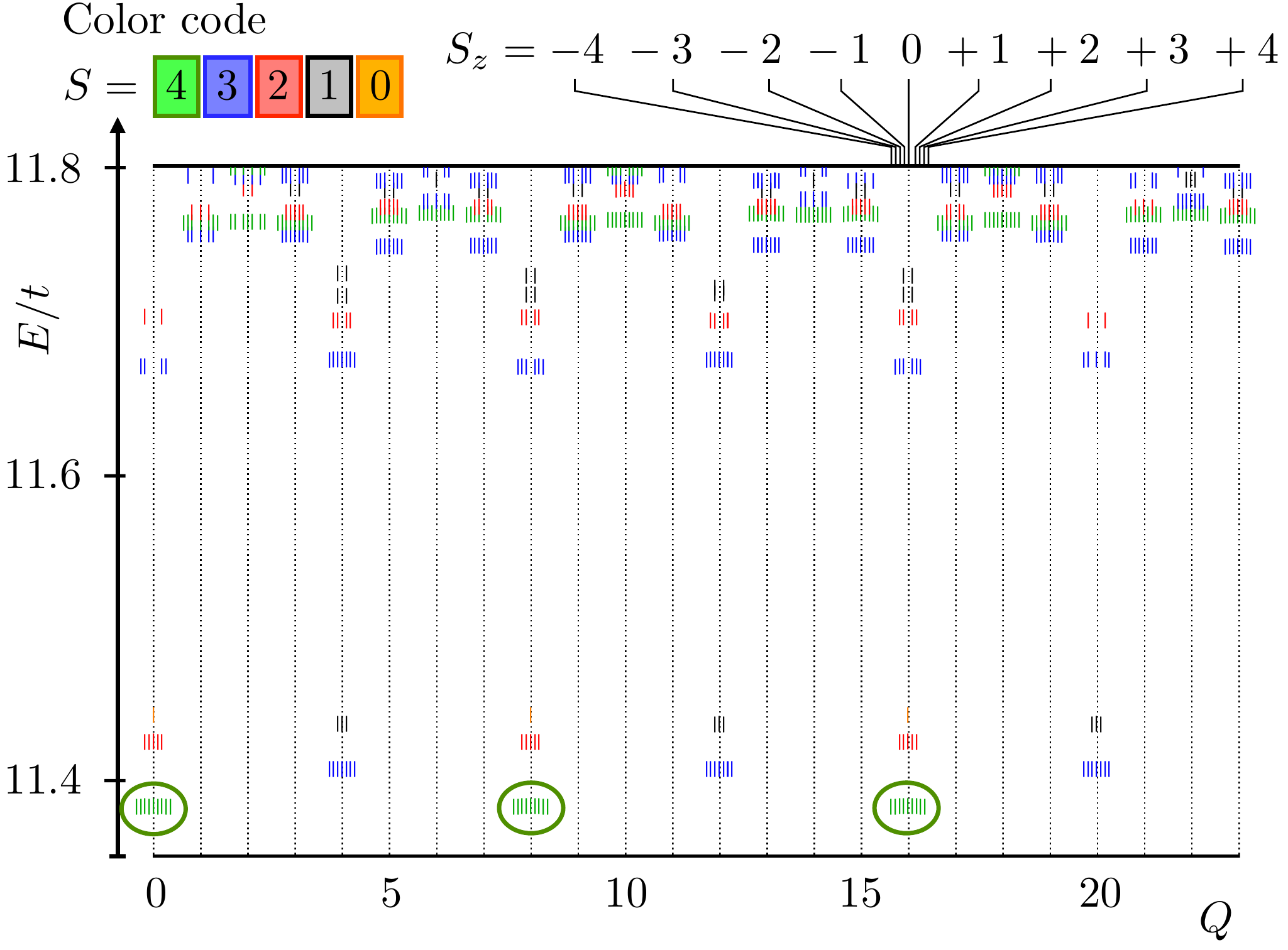}
\caption{\label{Fig1}Eigenvalue spectrum as obtained with exact diagonalization of $\hat{H}_{\text{eff}}$ in \eqref{eq:tauindep}
for a $L_x \times L_{y} = 4\times 6$ lattice. The parameters are the same as in Ref.~\cite{GGN14}, Fig 3 except that we choose a lower amplitude 
($A_{x}=A_{y}=1.0$), rending our proposal even more viable.}
\end{figure}
In contrast to $\hat{H}_{\text{eff}}$, we note that $\hat{H}_{\text{F}}(\tau)$ includes to order $\omega^{-1}$ 
a) terms that depend explicitly on the initial time $\tau$ and 
b) interaction corrections of the form discussed in Ref.~\cite{D14} through the commutators $[\hat{H}_{\pm n}, \hat{H}_0]$. 
Both a) and b) terms generically break point group symmetries of the lattice. Importantly, $\hat{H}_{\text{F}}(\tau)$ and $\hat{H}_{\text{eff}}$ are unitarily equivalent (if all orders of $\omega$ are included)
\begin{equation}
\label{eq:hams}
\hat{H}_{\text{F}}(\tau) = e^{i K(\tau)} \hat{H}_\text{eff} e^{-i K(\tau)} .
\end{equation}
As a consequence, the spectra of both Hamiltonians are identical. \\

\begin{figure}
\includegraphics[scale=0.42]{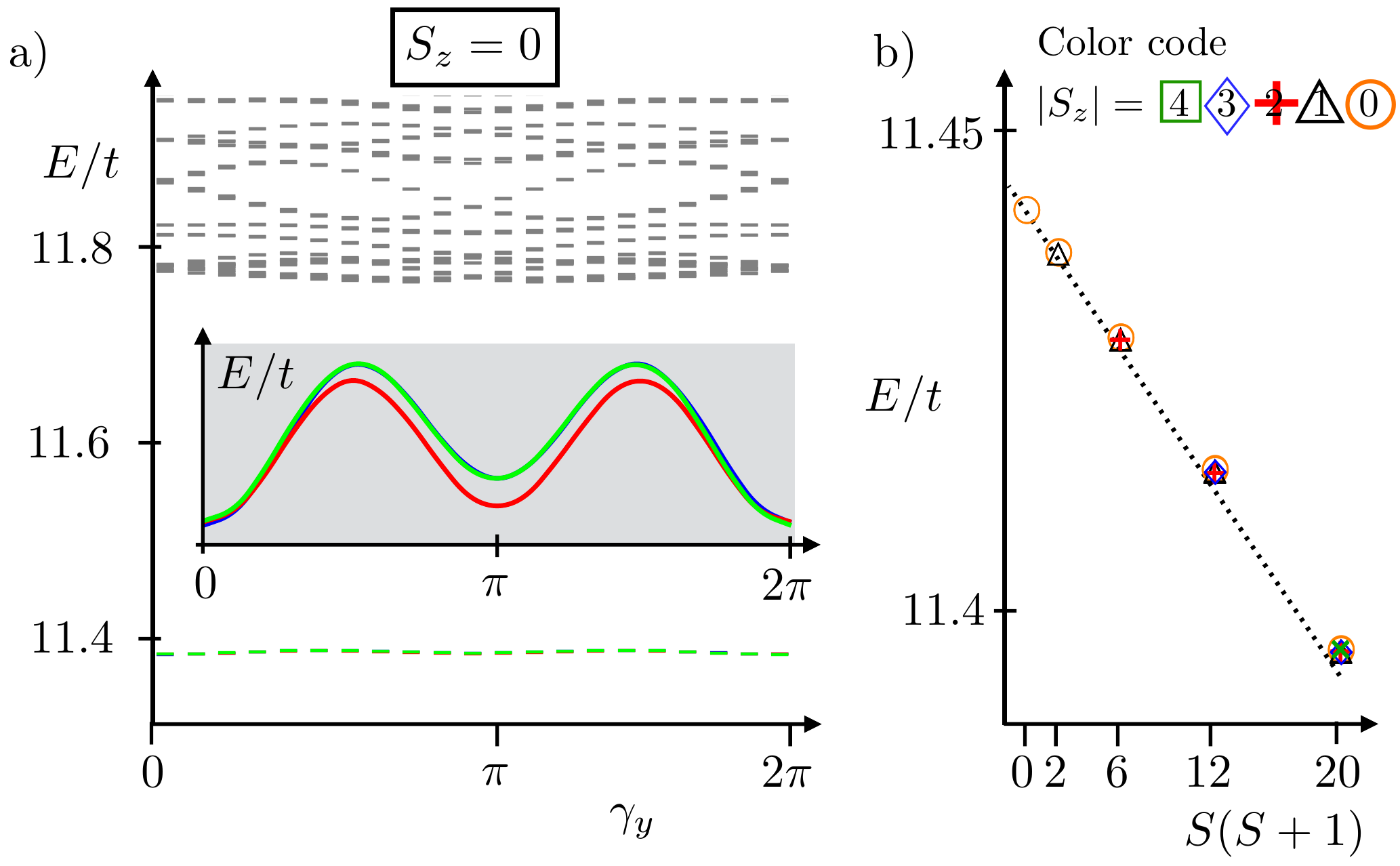}
\caption{\label{Fig2} a) Flux insertion and b) tower of states as explained in Ref.~\cite{GGN14} evidencing the fractional and ferromagnetic nature of the ground state, respectively. 
Note that the spectrum in a) is symmetric around $\phi=\pi$ which is not the case when $e^{\pm n\omega\tau}[H_{0},H_{\pm n}]$ commutators are included. The inset in a) shows how the three lowest states permute under flux insertion.}
\end{figure}

The universal time-evolution of the system, independent of initial conditions, is thus determined by $\hat{H}_{\text{eff}}$.  
The central question that was addressed in our Letter Ref.~\cite{GGN14}
is, whether this time-evolution operator can support a topologically ordered Floquet fractional Chern insulator (FFCI) as a steady state of the system. The difference between the effective Hamiltonian used to obtain the results in Ref.~\cite{GGN14} and $\hat{H}_{\text{eff}}$ are that those terms of $[\hat{H}_{\pm n}, \hat{H}_0]$ that add to the non-interacting Hamiltonian were erroneously included in Ref.~\cite{GGN14}, while they should not be contained in a consistent approximation to the effective Hamiltonian to order $1/\omega$.
However, the presence or absence of these terms does not affect qualitatively the main result of Ref.~\cite{GGN14}, namely an affirmative answer to the above question. It was pointed out in Ref.~\cite{Eckardt2015} that, even though the terms $[\hat{H}_{\pm n}, \hat{H}_0]$ appear at order $1/\omega$ in the Floquet-Magnus expansion, they affect the spectrum only at order $1/\omega^2$. As a numerical confirmation, we have exactly diagonalized $\hat{H}_{\text{eff}}$ as given by Eq.~\eqref{eq:tauindep} with the terms up to order $\omega^{-1}$, i.e. without the $\tau$-dependent terms. We found that the nature of the ground state being an FFCI with spontaneous breaking of spin-rotation symmetry is unaffected by this change (see Fig.~\ref{Fig1} and ~\ref{Fig2}). In fact, the gap above the FFCI ground states in a given spin sector is found to be larger than in Ref.~\cite{GGN14} and the FFCI phase is stable at lower amplitudes of the driving field.
 
This numerical result does not answer (and did not attempt to answer) the question
 of how to initialize the system through the kick operator $K(\tau)$ in order to obtain this desired steady state .
It is the power of the decomposition \eqref{eq:evol} that allows to separate the two problems. In contrast, were we to use $\hat{H}_{\text{F}}(\tau)$ as Ref.~\cite{D14} advocates, the effect of the kick operators and the time evolution are entangled, and thus less transparent to work with.

This leads us to statement ii) from Ref.~\cite{D14}: How to reach and stabilize the desired FFCI steady state. The answer to this question lies in a deliberate control over the initialization of the system as well as its coupling to a heat bath and goes beyond the scope of Ref.~\cite{GGN14}. For example, if the bath density of states is suppressed at energies that correspond to unwanted transitions, a thermal-like population of Floquet could be possible.~\cite{DOM14a,DOM14b,Seetharam15,Iadecola2015}
Once reached, the transitions out of the FFCI state may be suppressed for exponentially long times in a limit where the driving frequency is large compared to all characteristic energy scales of the undriven system, even in absence of a bath. \footnote{D. Abanin, private communication.}

Given the above discussion, we conclude that a) there is no correction to order $\omega^{-1}$ in the interaction when calculating $\hat{H}_{\text{eff}}$, b) this Hamiltonian hosts an FFCI state and c) reaching this steady state requires deliberate control over the initialization of the system and its coupling to a bath. In summary, while Refs.\cite{BDP14} and~\cite{D14} raised important points about the derivation of effective Hamiltonians in time-periodic systems and their relation to the actual steady states, we can confirm that our main result is unaffected.

\textit{Acknowledgements - } We are grateful to D. Abanin, C. Chamon, A. Eckardt, T. Iadecola, and T. Oka for
illuminating discussions. We wish to acknowledge as well discussions with 
M. Bukov and L. D'Alessio on the origin of the apparent controversy.

%\bibliography{commFCI.bib}

\end{document}